\documentclass[twocolumn,twoside,slac_two]{revtex4}
\usepackage{graphicx}
\usepackage{fancyhdr}
\begin{document}

\title{Overview of the T2K long baseline neutrino oscillation experiment}


\author{Trung Le for T2K collaboration}
\affiliation{Department of Physics and Astronomy, Stony Brook University, NY 11794, USA}

\begin{abstract}
Neutrino oscillations were discovered by atmospheric and solar neutrino experiments, and have been confirmed by experiments using neutrinos from accelerators and nuclear reactors. It has been found that there are large mixing angles in the $\nu_e \rightarrow \nu_\mu$ and $\nu_\mu \rightarrow  \nu_\tau$ oscillations. The third mixing angle $\theta_{13}$, which parameterizes the mixing between the first and the third generation, is constrainted to be small by the CHOOZ experiment result. The T2K experiment is a long baseline neutrino oscillation experiment that uses intense neutrino beam produced at J-PARC and Super-Kamiokande detector at 295 km as the far detector to measure $\theta_{13}$ using $\nu_e$ appearance. In this talk, we will give an overview of the experiment.
\end{abstract}

\maketitle

\thispagestyle{fancy}


\section{Introduction}
In recent years, major progress has been made in neutrino physics, especially with regard to neutrino masses and neutrino oscillations. Neutrino oscillations were discovered in atmospheric\cite{atm} and solar neutrinos\cite{solar,sno}, and they have been confirmed by man-made neutrinos produced in accelerators\cite{k2k, minos} and nuclear reactors\cite{kamland}. Atmospheric and accelerator neutrino oscillation experiments measure the mixing angle $\theta_{23}$ which parameterizes the mixing of the second and the third lepton generation and the corresponding squared mass difference $\Delta m^2_{23}$. Solar and reactor (with baseline around 100 km) neutrino oscillation experiments measure the mixing angle $\theta_{12}$ between the first and second generation. The correct sign of $\Delta m^2_{21}$ is established thanks to the matter effects in solar neutrino oscillation. The neutrino oscillation results require that neutrinos have masses and there are at least three distinct masses. However, the neutrino oscillation between the first and the third generation has not been found. Currently there is a limit on the mixing angle $\theta_{13}$ from the CHOOZ reactor experiment\cite{chooz}. It is interesting to see if this mixing angle is nonzero.

T2K is a second generation long baseline neutrino oscillation experiment to measure oscillation parameters, especially the mixing angle $\theta_{13}$ through  $\nu_e$ appearance from a $\nu_\mu$ beam. The T2K $\nu_\mu$ beam is generated using the high intensity 50 GeV proton synchrotron at J-PARC in Tokai, Japan and the far detector is Super-Kamiokande which is located 295 km from the accelerator. The near detector is being built and installed 280 m from the target. We will briefly review the theory of neutrino oscillations in the next section and describe various components of the experiment in the following sections.
\section { Three-flavor neutrino oscillations}
Neutrino mixing is a direct consequence of neutrino masses. In this case the weak  eigenstates $\nu_f = (\nu_e,\nu_\mu,\nu_\tau)$, i.e., states that are produced from weak decays, are not the same as the mass eigenstates $\nu_m=(\nu_1,\nu_2,\nu_3)$, but instead are linear combination of the mass eigenstates,
\begin{equation}
\nu_f = U \nu_m,
\end{equation}
where $U$ is the mixing matrix. The mixing matrix is usually parameterized using three mixing angles $(\theta_{12}, \theta_{23},\theta_{13})$ and one CP violation phase $\delta$\cite{keung}
\begin{eqnarray}
U&=&\left(
\begin{array}{ccc}
1 & 0 & 0 \\
0 & c_{23} & s_{23}\\
0 & -s_{23} & c_{23}
\end{array}
\right) \left(
\begin{array}{ccc}
c_{13} & 0 & s_{13} e^{-i\delta} \\
0 & 1 & 0 \\
-s_{13}e^{i\delta} & 0 & c_{13}
\end{array}
\right) \times \nonumber \\ & &\left(
\begin{array}{ccc}
c_{12} & s_{12} & 0 \\
-s_{12} & c_{12} & 0 \\
0 & 0 & 1
\end{array}
\right),
\end{eqnarray}
where  $s_{ij} \equiv \sin\theta_{ij}$ and $c_{ij} \equiv \cos\theta_{ij}$. The oscillation probabilities in case of three flavor neutrinos obtained in the quantum mechanical treatment of neutrino oscillations are
\begin{eqnarray}\label{neuprob}
P(\nu_\alpha \rightarrow \nu_\beta)=\delta_{\alpha \beta}&-&4\sum_{i>j}\text{Re}J^{\alpha \beta}_{ij}\sin^2
\phi_{ij}\\ \nonumber&+&2 \sum_{i>j}\text{Im}J^{\alpha \beta}_{ij}\sin 2 \phi_{ij},
\end{eqnarray}
where $\phi_{ij}$ is the oscillation phase and
\begin{equation}
J^{\alpha \beta}_{ij}=U_{\alpha i}U_{\alpha j}^* U_{\beta i}^* U_{\beta j}.
\end{equation}
Of the oscillation channels, the $\nu_\mu \rightarrow \nu_e$ appearance channel is of particular interest. The full oscillation probability is complicated, but when neglecting the CP violation terms and matter effects, the oscillation probability becomes\cite{martin}
\begin{equation}
P(\nu_\mu \rightarrow \nu_e) = \sin^2 \theta_{23} \sin^22\theta_{13}\sin^2\frac{\Delta m_{32}^2 L}{4E_\nu}.
\label{appearance}
\end{equation}
It is noticed that the amplitude of the oscillation is proportional to $\sin^22 \theta_{13}$. The measurement of this oscillation channel will give a direct measurement of the mixing angle $\theta_{13}$.

\section {T2K physics goals}
The T2K experiment aims to measure the mixing angle $\theta_{13}$ through $\nu_e$ appearance. It is noted that the appearance probability (\ref{appearance}) also depends on the atmospheric oscillation parameters  ($\Delta m_{23}^2, \sin^22\theta_{23}$). For this reason, the T2K experiment will measure both these oscillation parameters and the mixing angle $\theta_{13}$.

{\bf $\nu_\mu$ disappearance} The oscillation parameters ($\Delta m_{23}^2, \sin^22\theta_{23}$) will be determined from the survival probability of $\nu_\mu$ after traveling 295 km. This probability can be measured by comparing the neutrino spectra at the near and far site. The neutrino spectrum at the near site is measured by the near detector system. The neutrino flux is also measured at the near site and extrapolated to the far site to obtain the correct rate normalization. At the far site, the neutrino spectrum is measured by the Super-Kamiokande detector. Muon neutrinos are detected at Super-Kamiokande using the quasi-elastic charged current interaction which has better energy reconstruction. The muons are identified by the presence of a muon-like ring.

\begin{figure}[h]
\centering
\includegraphics[width=80mm]{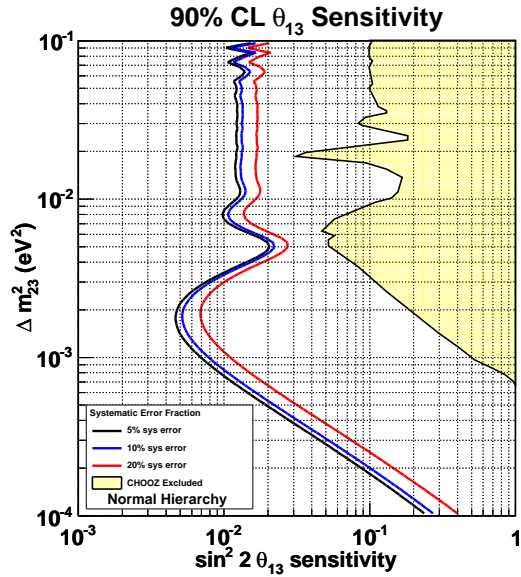}
\caption{T2K sensitivity to $\theta_{13}$ at the 90\% confidence level as a function of $\Delta m^{2}_{23}$.  Beam is assumed to be running at 750kW for 5 years (or equivalently, 5$\times 10^{21}$ POT), using the 22.5 kton fiducial volume SK detector.  5\%, 10\% and 20\% systematic error fractions are plotted.  The yellow region has already been excluded to 90\% confidence level by the CHOOZ reactor experiment.  The following oscillation
parameters are assumed: $\sin^{2}2\theta_{12} = 0.8704$, $\sin^{2}2\theta_{23} = 1.0$, $\Delta m^{2}_{12} =  7.6\times10^{-5}\text{eV}^{2}$, $\delta_{CP}=0$, normal hierarchy.} 
\label{sensitivity}
\end{figure}

{\bf $\nu_e$ appearance} The mixing angle $\theta_{13}$ is determined from the measurement of the $\nu_e$ appearance signal. The $\nu_e$ signal comes from $\nu_\mu \rightarrow \nu_e$ oscillation if the mixing angle $\theta_{13}$ is nonzero. Electrons from $\nu_e$ quasi-elastic charged-current interactions are detected by the presence of a electron-like ring. Backgrounds from $\nu_\mu$ interactions are further reduced by requiring that there is no decay electron. After this, there are two main backgrounds to the $\nu_e$ search at Super-Kamiokande:  the background from single $\pi^0$ from neutral current interactions and $\nu_e$ intrinsic to the beam. These backgrounds are measured by the near detector system before oscillation and extrapolated to Super-Kamiokande. The estimated $\pi^0$ and intrinsic $\nu_e$ backgrounds  are subtracted from the selected $\nu_e$ events. The final appearance signal is fitted to the appearance probability (\ref{appearance}) to find $\sin^22\theta_{13}$. Fig.~\ref{sensitivity} shows T2K sensitivity to $\theta_{13}$ at 90\% confidence level as a function of $\Delta m^2_{23}$, assuming no CP violation ($\delta_{CP}$ = 0) and normal mass hierarchy.

\section { Neutrino beam and monitor}
The neutrino beam is produced by smashing protons from the J-PARC proton synchrotron on a target. The main synchrotron is designed to accelerate protons up to 50 GeV, however, the initial proton energy is limited to 30 GeV. The proton beam is extracted by the neutrino primary beamline. The neutrino beamline consists of 28 combined function superconducting magnets which produce both dipole and quadrapole fields and normal magnets near the final focusing sections. The design intensity of the proton beam is $3.3\times 10^{14}$ at the rate of about $0.3$ Hz. The target is a graphite cylinder of 30 mm in diameter and 900 mm in length. Three electromagnetic horns are used to focus (and select the right charge sign) charged pions generated in the target to the forward direction. The target is installed inside the inner conductor of the first horn to collect and focus the pions as much as possible. These horns are driven by a pulsed current of 320 kA synchronized with the beam. The focused pions decay into $\nu_\mu$ and muons in a 110 m decay tunnel which follows the horns. There is a small fraction of $\nu_e$ (about .5\% at peak energy) produced by decaying muons and kaons. The decay tunnel is filled with 1 atm helium gas to reduce pion absorption and tritium production. Charged particles are stopped by the beam dump placed at the end of the decay tunnel. The beam dump is designed to allow high energy muons ($>$ 5 GeV) passing through. These muons are used to monitor the neutrino beam.
\begin{figure}[h]
\centering
\includegraphics[width=80mm]{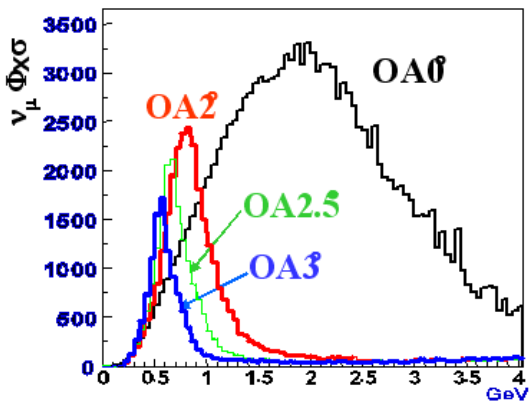}
\caption{Neutrino energy spectra at SK at different off-axis angles, 3$^0$, 2.5$^0$, 2$^0$ and on-axis.} 
\label{spectrum}
\end{figure}

\begin{figure}[h]
\centering
\includegraphics[width=80mm]{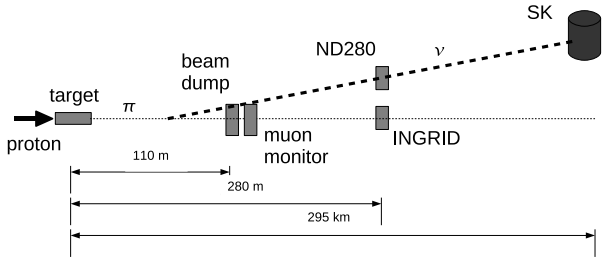}
\caption{Schematic of the T2K neutrino beam}
\label{schematic}
\end{figure}

The T2K neutrino beamline adopts an off-axis beam configuration\cite{beavis}.  It exploits the kinematics of pion decay that the neutrino energy is not strongly dependent on the pion energy at a fixed decay angle in the lab frame to produce a narrow-band beam (Fig.~\ref{spectrum}). The narrow-band beam is desired to maximize the neutrino flux at energies near the first oscillation maximum. The off-axis angle can be changed from 2.0 to 3.0 degrees. This corresponds to the mean neutrino energies from 0.5 to 0.9 GeV. The nominal off-axis angle is 2.5 degrees, corresponding to the peak beam energy of about 0.7 GeV. The schematic of the T2K neutrino beam is shown in Fig.~\ref{schematic}.

Because of the off-axis beam configuration, the neutrino spectrum at Super-Kamiokande is sensitive to the neutrino beam direction. For this reason there are two detector systems designed specifically for online neutrino beam monitoring: One is a muon monitor and the other is an on-axis neutrino detector. The muon monitor can measure the neutrino beam condition in real time by detecting the accompanying muons. The on-axis neutrino detector monitors the neutrino beam directly using neutrino interactions. The muon monitor and the on-axis neutrino detector are described below.

\subsection { Muon monitor (MUMON)} The muon monitor is placed downstream of the beam dump and monitors the direction, profile, time structure,  and intensity of the beam by detecting high energy muons which are produced together with the neutrinos. Thanks to the high muon flux, the MUMON is sensitive to the proton hit position on the target and the status of the target and horns. Therefore, it is also used as a proton beam monitor, a target monitor, and a horn monitor. The measurements of MUMON can monitor the quality of the neutrino beam on a spill-by-spill basis. Finally, the MUMON helps aim the neutrino beam at Super-Kamiokande during the beginning of the experiment. The MUMON consists of two independent detectors: a matrix of silicon detectors and an array segmented ionization chambers. Each detector covers an area of 1.5m $\times$ 1.5m. The silicon detector matrix consists of 7 $\times $ 7 silicon photodiodes mounted on the upstream panel. The silicon photodiodes are not radiation hard and can only be used in the early stage of the experiment. More radiation tolerant detectors like diamond detector are being tested. The ionization chamber detector consists of an array of 7 segmented ionization chambers on the downstream side.

\subsection { On-axis detector (INGRID) } 
\begin{figure}[h]
\centering
\includegraphics[width=80mm]{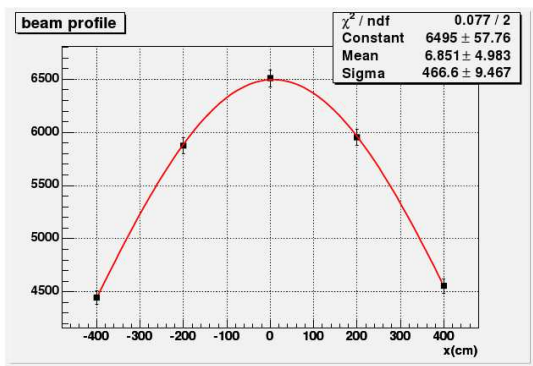}
\caption{Neutrino beam profile measured with the INGRID detector (Monte Carlo simulation).} 
\label{profile}
\end{figure}

The INGRID detector is located on-axis at 280 m from the target and beneath the off-axis detector. It monitors the neutrino beam by using muons from charged current neutrino interactions. Because of the small neutrino cross sections, it can only monitor the neutrino beam on a daily basis. The total number of neutrino events observed by INGRID is about 10,000 events/day. The detector consists of 16 modules arranged in 7 vertical and 7 horizontal modules and two off-axis modules. Each module has dimensions of 1.2m $\times $ 1.2m $\times$ 1.3m and consists of 11 scintillator planes alternating with 10 iron targets. On the top and sides of each module, three or four additional scintillator layers are used as a veto. Each tracking plane has 24 rectangular scintillator bars, each of dimensions of 5cm$\times$120.3cm$\times$1cm.  Fig.~\ref{profile} shows the neutrino beam profile measured from the INGRID detector (Monte Carlo simulation).

\begin{figure*}[hbt]
\centering
\includegraphics[width=120mm]{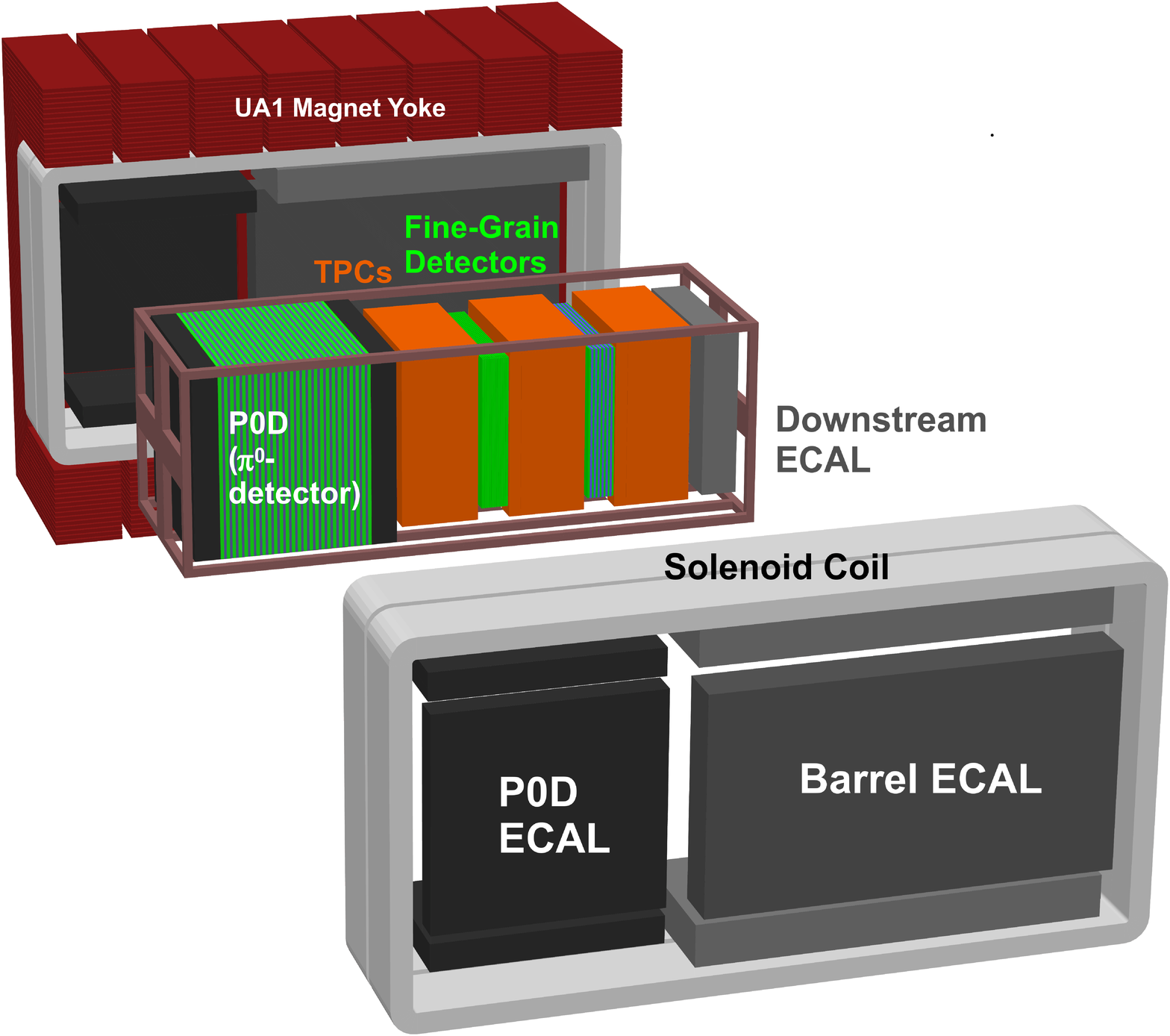}
\caption{The off-axis near detector system shown with one side of the UA1 magnet. The inner detectors are supported by a basket and consist of the P0D upstream, followed by the tracker, and the dowstream ECAL. They are surrounded by the side ECALs.} \label{nd280}
\end{figure*}

\section { Near detector system (ND280)}
The near detector system is located off-axis at 280 m from the target and consists of different sub-detectors. The main purpose of the near detector system is to measure the properties of the neutrino beam before oscillation:
\begin{itemize}
  \item Measure the neutrino flux and spectrum
  \item Measure different interaction cross sections to estimate the backgrounds at Super-Kamiokande.
  \item Measure the $\nu_e$ beam contamination for $\nu_e$ appearance search.
\end{itemize}
The ND280 sub-detectors are enclosed inside the UA1 dipole magnet operating at nominal 0.2 T (Fig.~\ref{nd280}). The magnetic field is used to reduce electron diffusion inside the time projection chambers and bend charged particle trajectories for momentum measurement. The sub-detectors are described below.
\subsection { Pizero detector (P0D) }
One of the dominant backgrounds to the $\nu_e$ appearance search in water Cherenkov detector is single $\pi^0$ from neutral current interactions. The P0D is designed to measure the neutral current single $\pi^0$ production cross section on water. Using this cross sections, the $\pi^0$ background at SK can be estimated.  Specifically, it will measure the neutrino interactions with water in and water out, and events measured with water out will be statistically substracted from the events with water in to estimate the event rate on water. 

The P0D consists of a water target sandwiched between two electromagnetic calorimeters (ECAL).  Note that these ECAL sections are part of the P0D, and not the ECALs mentioned in the following sections. The water target section consists of 26 tracking modules alternating with water modules. Each water module has two water bags supported by high-density polyethylene frame. Radiator made of 2 mm brass sheet is inserted between tracking module and water module to increase the effective radiation length. Each ECAL section, at the front and back of the detector, has 7 tracking modules alternating with 4 mm thick lead sheet. Each tracking module is a complete tracking unit of dimensions of 220cm$\times$230cm$\times$3.9cm, it has x-y tracking planes perpendicular to the beam direction. Each tracking plane is made of 134 for $y$ direction (126 for $x$ direction) triangular plastic scintillator bars, each of 1.7 cm height and 3.4 cm base and has a axial hole at the center. Scintillation light is collected by a Kuraray Y-11 wavelength-shifting fiber inserted in the hole. The signal is readout at one end by a Multi-Pixel Photon Counter (MPPC) developed by Hamamatsu Photonics\cite{mppc}, the other end of the fiber is mirrored. On the opposite side of the MPPCs there are two back-to-back LEDs used to inject UV light into the wavelength-shifting fibers for calibration purpose.

\subsection {Tracker: Fine-grained detector and time projection chamber}
The other dominant background to the $\nu_e$ appearance search is the $\nu_e$ beam contamination which is about 0.5\% of the $\nu_\mu$ flux at peak energy. This background is resulted from kaon and muon decays, it can not be removed from the $\nu_\mu$ beam and must be measured by the near detector. One of the purpose of the tracker is to measure this background. The tracker also measures $\nu_\mu$ flux and spectrum before oscillation for $\nu_\mu$ disapperance study. 

Finally, in addition to the flux and spectrum measurement, the tracker can distinguish the simple quasi-elastic charged current interaction from non-elastic interactions. In the fine-grained detector, this is accomplished by the fine segmentation which allows tracking of low energy protons. The presence of both proton and muon tracks create a kinematical constraint to remove non-elastic events. Furthermore, the good particle identification of the time projection chambers using ionization energy loss can distinguish electrons, muons, and protons. The time projection chambers can also measure the charge sign of charged particles to further reject non-elastic events.
%Since oscillation analyses are done using the simple quasi-elastic charged current interaction, the fine-grained detector can track low energy proton. This kinematic constraint can be used to distinguish CCQE from non-elastic interactions. In addition to neutrino spectrum measurement, the detector must have good particle identification capability. The charge sign of the neutrino interaction products must be determined unambiguously in order to distinguish charged pions produced in inelastic interactions and the anti-neutrino component of the beam. Good measurements of ionization energy loss are needed to distinguish electrons, muons, and protons.

\paragraph {Fine-grained detector (FGD)} There are two FGDs, each of dimensions of 200cm $\times$ 200cm $\times$ 30cm. One FGD consists of 30 tracking planes. These tracking planes are arranged in alternating vertical and horizontal direction perpendicular to the beam. The back FGD consists of alternating x-y tracking planes and 3 cm thick layers of water target. For both FGDs, each scintillator plane consists of 200 scintillator bars, each of dimensions 1.0cm $\times$ 1.0cm $\times$ 200cm, has a central hole for wavelength shifting fiber and TiO$_2$ coating. The fine-grained segmentation allows tracking of low energy protons to distinguish CCQE and non-elastic events. Light collected by the wavelength-shifting fiber is read out by a MPPC.
\paragraph {Time projection chamber (TPC)} 
The TPCs are optimized to measure charged particles from neutrino interaction in the FGDs and the P0D. There are three TPCs sandwiching with the FGDs, with one TPC downstream of the P0D. Each TPC module has a dimensions of 180cm $\times$ 200cm $\times$ 70cm (sensitive volume). It has a double wall structure, the inner wall makes up the field cage and the outer wall is used for gas, high voltage insulation. The sensitive volume contains a mixture of gases Ar-CF$_4$-iC$_4$H$_{10}$ (95\%-3\%-2\%) and has drifting velocity of 7.8cm/$\mu s$ at 280 V/cm. Gas amplification and readout using Micromegas with pad size of 7mm $\times$ 10mm.

\subsection { Electromagnetic calorimeter (ECAL)}
Electromagnetic calorimeters surround the P0D and the tracker to detect showering ($e^-,\gamma$) particles from neutrino interactions in these detectors. Charged particles are produced and tracked by the inner detectors. However, because of the low mass of the inner detectors, showering particles can escape and cause energy leakage which reduces the energy resolution. The ECALs are designed to have a short effective radiation length to convert these particles. Showers found in the ECALs are matched to tracks or showers found in the inner detectors. The shower-to-track matching helps distinguish muon from electron while the shower-to-shower matching reduces the energy leakage. Furthermore, the ECALs also improves the $\pi^0$ detection efficiency by increasing the probability of catching the $\gamma$'s from $\pi^0$ decay. Finally, the ECALs also acts as a veto detector to detect particles from neutrino interactions in the magnet.The ECALs consist of alternating layers of plastic scintillator and lead.

\subsection {Side muon range detector (SMRD)}
It is important to measure muon at high angle relative to the beam direction to increase the acceptance. As the name implies, the SMRD measures muon range from which the momentum can be estimated. The SMRD also acts as the veto detector for cosmic ray muons and is used to form the cosmic trigger. The SMRD is constructed by inserting scintillator detectors into the gaps between iron plates of the magnet. 

\section { Far detector - Super-Kamiokande}
The far detector Super-Kamiokande (SK) is located 295 km from the near detector at the Kamioka Observatory, Gifu, Japan. Detailed description and analysis of the detector can be found in \cite{sk1,sk2}. A brief description is given here. SK is a 50 kton water Cherenkov detector. The detector is a cylindrical tank of 41.4 m in height and 39.3 m in diameter. The tank is optically divided into an inner detector of 36.2 m in height and 33.8 m in diameter and an outer detector. Both sides of the dividing wall are mounted with photomultipliers (PMTs), 11146 20-inch diameter PMTs on the inside facing inward and 1885 8-inch diameter PMTs on the outside facing outward. Cherenkov light from particles are recorded by these PMTs. The detector can distinguish electron from muon by looking at the light distribution of the projected Cherenkov cone. Electrons scatter more than muons, and therefore make a fuzzy ring.

It is well-known that water Cherenkov detector can not distinguish between gamma and electron, they both produce e-like ring. Because of this, SK could mistake $\pi^0$ for electron when the two gamma's from the $\pi^0$ decay have a small open angle or large energy asymmetry. In the case of small open angle, the two gamma's look like a single gamma and in the case of large energy asymmetry, the smaller energy gamma is not detected. 

\begin{figure}[htb]
\centering
\includegraphics[width=80mm]{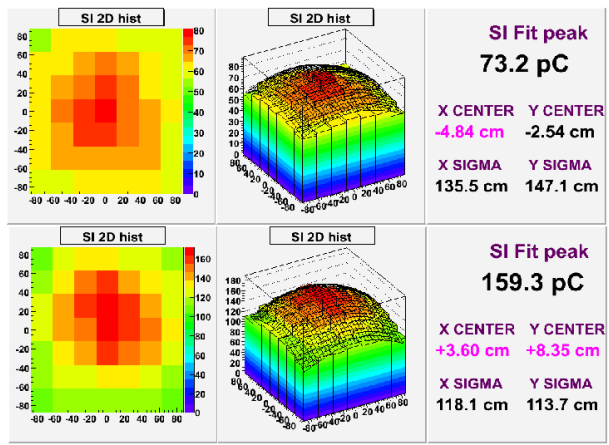}
\caption{Muon monitor profile from the MUMON silicon detector with horn on (bottom) and off (top). } 
\label{mumon}
\end{figure}

\section { Status of the experiment }
The neutrino beam was commissioned in April, 2009. Muons which are produced together with neutrinos from pion decays were detected by the MUMON. The muon monitor profile from the silicon detector with the horn on and off is shown in Fig.~\ref{mumon}. The peak and standard deviation of the muon distributions show that the horn is working as expected. The construction and installation of the INGRID detector are finished, and it has been taking cosmic data. All the P0D modules were tested with cosmic ray muons, the P0D installation has been completed. The tracker (TPC + FGD) construction is done (one TPC will be installed later), it went through test beam at TRIUMF and currently awaits to be installed. The downstream ECAL was tested with test beam at CERN. The ECALs around the tracker and the P0D will be installed next year. Finally, the SK detector is ready for neutrino beam.
\\
\section { Conclusion}
The T2K experiment is a long baseline neutrino oscillation experiment to measure the last mixing angle $\theta_{13}$ using the $\nu_e \rightarrow \nu_\mu$ appearance channel. The experiment uses the neutrino beam produced at J-PARC which has full power of 0.75 MW and the Super-Kamioakande as the far detector at 295 km away. The near detector system monitors and characterizes the neutrino beam before oscillation. The neutrino beam has been commissioned and the near detector installation is being completed. The experiment will start taking data in December 2009.

%\section{Paper Submission}

%Authors should submit their papers to the ePrint arXiv 
%server\footnote{http://arxiv.org/help} 
%after verifying that it is processed correctly by the LaTeX processor.
%Please submit the source code, the style files 
%(revsymb.sty, revtex4.cls, slac\_two.rtx) 
%and any figures; 
%these should be self-contained to generate the paper from source.  

%It is the author's responsibility to ensure that the papers are 
%generated correctly from the source code at the ePrint server. 
%After the paper is accepted by the ePrint server, please verify that
%the layout in the resulting  PDF file conforms to the guidelines 
%described in this document.
%Finally, contact the organizers of DPF-2009 and your parallel session conveners 
%(see http://www.dpf2009.wayne.edu) with the ePrint number of the paper; 
%the deadline to do this is 2~October~2009.

%\begin{acknowledgments}

%\end{acknowledgments}

\bigskip % extra skip inserted

\end{document}